\documentclass[aps,prl,twocolumn,superscriptaddress,showpacs]{revtex4}
\usepackage{graphicx}
\usepackage{dcolumn}
\usepackage{bm}
\usepackage{color}

\newcommand{\Gam}{\Gamma}
\newcommand{\ew}[1]{\left\langle #1 \right\rangle}
\newcommand{\mr}[1]{\mathrm{#1}}

\newcommand{\om}{\omega}
\newcommand{\vp}{\varphi}
\newcommand{\RE}{\,\mathrm{Re}\,}

\begin{document}
\title{Dynamical Coulomb Blockade Observed in Nano-Sized Electrical Contacts}

\author{Christophe Brun}
\affiliation{Institut de Physique de la Mati\`ere Condens\'ee, Ecole Polytechnique F\'ed\'erale de Lausanne, CH-1015 Lausanne, Switzerland}
\affiliation{Institut des Nanosciences de Paris, CNRS-UMR 7588, Universit\'e Pierre et Marie Curie-Paris 6 UPMC, F-75252, Paris, France}
\author{Konrad H. M\"uller}
\affiliation{D\'epartement de Physique Th\'eorique, Universit\'e de Gen\`eve, CH-1211 Gen\`eve, Switzerland}
\author{I-Po Hong}
\affiliation{Institut de Physique de la Mati\`ere Condens\'ee, Ecole Polytechnique F\'ed\'erale de Lausanne, CH-1015 Lausanne, Switzerland}
\affiliation{Institut f\"{u}r Experimentelle und Angewandte Physik, Christian-Albrechts-Universit\"{a}t zu Kiel, D-24098 Kiel, Germany}
\author{Fran\c{c}ois Patthey}
\affiliation{Institut de Physique de la Mati\`ere Condens\'ee, Ecole Polytechnique F\'ed\'erale de Lausanne, CH-1015 Lausanne, Switzerland}
\author{Christian Flindt}
\affiliation{D\'epartement de Physique Th\'eorique, Universit\'e de Gen\`eve, CH-1211 Gen\`eve, Switzerland}
\author{Wolf-Dieter Schneider}
\affiliation{Institut de Physique de la Mati\`ere Condens\'ee, Ecole Polytechnique F\'ed\'erale de Lausanne, CH-1015 Lausanne, Switzerland}

\date{\today}

\begin{abstract}
Electrical contacts between nano-engineered systems are expected to constitute the basic building blocks of future nano-scale electronics. However, the accurate characterization and understanding of electrical contacts at the nano-scale is an experimentally challenging task.  Here we employ low-temperature scanning tunneling spectroscopy to investigate the conductance of individual nano-contacts formed between flat Pb islands and their supporting substrates. We observe a suppression of the differential tunnel conductance at small bias voltages due to dynamical Coulomb blockade effects. The differential conductance spectra allow us to determine the capacitances and resistances of the electrical contacts which depend systematically on the island--substrate contact area. Calculations based on the theory of environmentally assisted tunneling agree well with the measurements.
\end{abstract}



\maketitle

\emph{Introduction.}---
The detailed understanding and precise engineering of electrical contacts between nano-sized metallic objects and their supporting substrates are not only of fundamental interest; they are also of crucial importance for the development of future electronics. As the nano-scale is approached, localization of charges in small regions of a device plays an increasingly important role and charging effects start to dominate the electron transport. Size reduction may for example lead to a change of the magnetic \cite{seneor2007} or superconducting \cite{bose2010} properties of a sub-micron contact. Investigations of electrical contacts in nano-scale systems are thus called for to help predict, design, and understand the functionality of nano-devices and thereby facilitate further progress in the fabrication of nano-scale electrical circuits~\cite{Lu2006,Likharev2003}.

One prominent charging mechanism is the dynamical Coulomb blockade (DCB) effect by which single electrons tunneling through a barrier exchange energy with the electromagnetic environment. These inelastic processes leave clear fingerprints in the current--voltage characteristics of a device and contain detailed information about the impedance of the electrical circuit in which the tunnel barrier is embedded.  The DCB effect was discussed theoretically in seminal works by Devoret \emph{et al.} \cite{Devoret1990}, Girvin  \emph{et al.} \cite{Girvin1990}, and Ingold and Nazarov \cite{Ingold1992}. Experimentally, it was observed in systems that were carefully engineered in order to obtain a high-impedance environment close to the tunnel barrier  \cite{Cleland1992,Holst1994,Joyez1997,Zheng1998,Pierre2001,Pekola2010,Saira2010,Parmentier2011}. The DCB effect, however, is not just an interesting physical phenomenon, it also leads to practical applications as we demonstrate below.

\begin{figure*}
\includegraphics[width=0.85\textwidth]{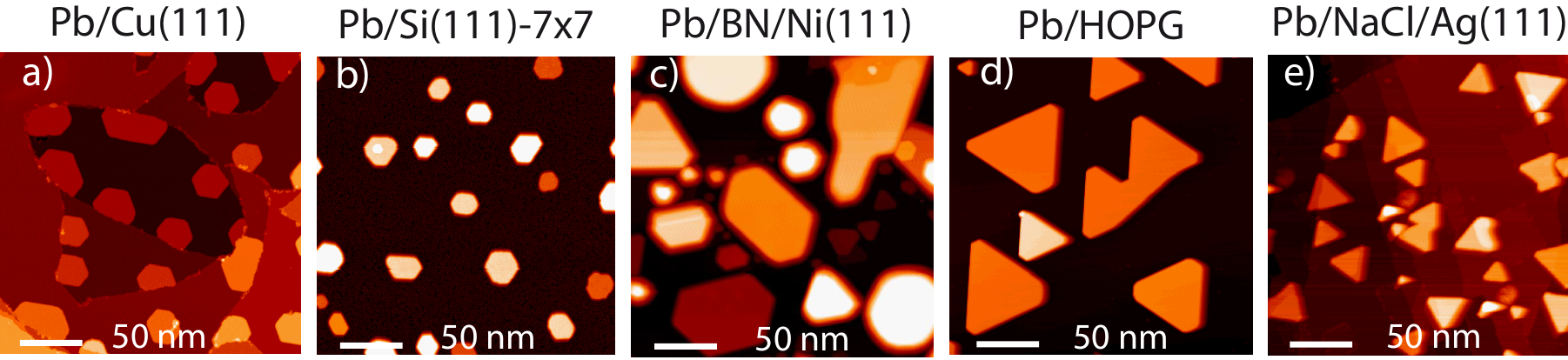}
\caption{\label{Pbislvarsubstrat}(color online). STM topographic images of flat Pb islands on different substrates. {\bf a)} metal Cu(111); island heights between
2 and 4 atomic monolayers (ML). {\bf b)} semiconductor Si(111)-7x7; 4-7 ML. {\bf c)} semi-metal HOPG; 7-10 ML. {\bf d)} metal covered by one insulating ML, h-BN/Ni(111); 6-60 ML. {\bf e)} metal covered by two insulating MLs, NaCl/Ag(111). The images were obtained with (from {\bf a} to {\bf e}) a voltage of 1.0~V, $-1.0$~V, $-0.5$~V, $-1.0$~V, and 3.0~V, and a corresponding tunnel current of 0.1~nA, $-0.1$~nA, $-0.1$~nA, $-0.05$~nA, and 0.02~nA.}
\end{figure*}

In this Letter we make use of DCB as a means to probe and characterize the electrical contact between flat metallic nano-scale islands and their supporting substrates. We employ low-temperature scanning tunneling spectroscopy to measure the electrical conductance of individual Pb islands on metallic, semimetallic, semiconducting, and partially insulating substrates. At low voltages we observe a suppression of the differential tunnel conductance due to DCB. Importantly, the tunnel current between the scanning tunneling microscope (STM) and the flat islands is highly sensitive to the impedance of the electrical contact between the island and the supporting substrate. This in turn allows us to investigate the nano-sized electrical contacts, employing the theory of DCB, and thereby extract the resistances and capacitances of the island--substrate contacts from the measured differential conductance spectra.

\emph{Experiment.}---
Figure~\ref{Pbislvarsubstrat} shows STM images of Pb(111) islands grown on substrates of Cu(111), Si(111)-7$\times$7, highly oriented pyrolytic graphite (HOPG), hexagonal (h-)BN/Ni(111), and NaCl/Ag(111). The island areas range from 10$\,$nm$^2$ to 10$^4\,$nm$^2$ with island heights between 2 and 60 monolayers (ML).  The substrate crystals were prepared according to standard procedures: The h-BN ML was epitaxially grown on Ni(111) \cite{Nagashima1995}. The Si(111) crystal was heavily $n$-doped and prepared to form a Si(111)-7$\times$7 reconstruction. NaCl was thermally evaporated onto Ag(111) at substrate temperatures between 300 K and 500 K \cite{Repp2006}. The flat islands were grown by evaporation of Pb from a W filament onto the substrates whose temperatures were stabilized between 130 K and 300 K to control the island sizes. We chose Pb as a typical metallic material because of the well-known growth of Pb films or islands on Si(111) \cite{Jalochowski1988,Weitering1992}, Cu(111) \cite{Hinch1989}, and HOPG \cite{Dil2007}. On HOPG, h-BN/Ni(111), and NaCl islands grow directly on top of the substrates, whereas on Si(111) and Cu(111) a 1 ML wetting layer forms first followed by the growth of single-crystal Pb islands \cite{Feng2004}.

Conduction measurements were performed in a home-built STM with PtIr tips operated under ultra-high vacuum at a temperature of $T=4.6$ K \cite{Gaisch1992}. We focused exclusively on islands supported by just a single substrate terrace such that the electrical contact between the island and the substrate was essentially uniform across the whole contact area. The differential conductance was measured in an open feedback loop using a lock-in technique with a peak--to--peak modulation voltage $V_{\rm pp}$ between $0.5$~mV and $5$~mV using a typical current of 1~nA and voltage ranges between $[-20,+20]$~mV and $[-200,+200]$~mV. We verified experimentally that the conductance spectra are independent of the injected power in a range between $10^{-11}$ and $10^{-9}$ W.

\begin{figure*}
\includegraphics[width=0.85\textwidth]{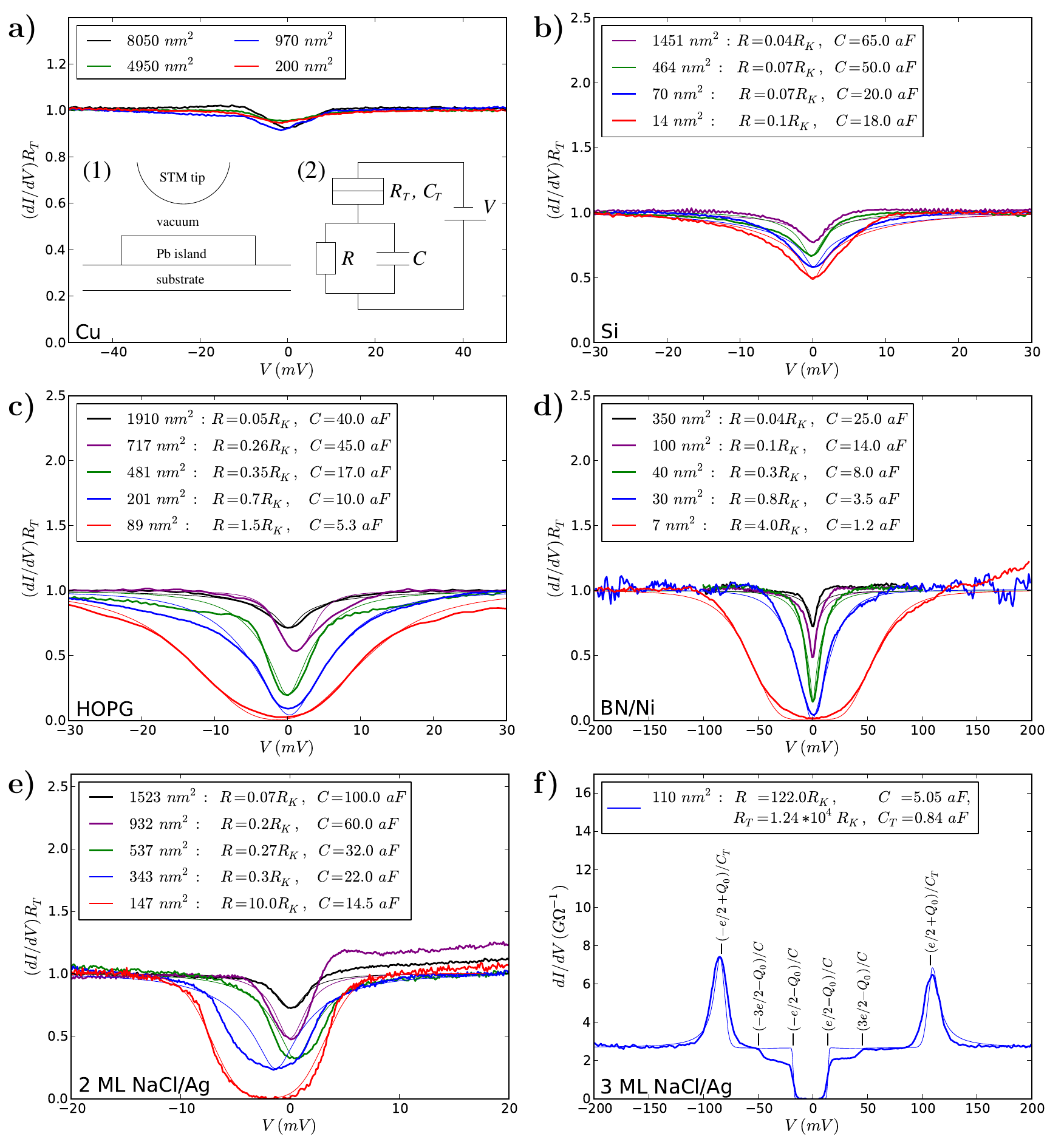}
\caption{\label{dIdVspectra} (color online). Differential tunnel conductance spectra. Experimental (thick lines) and theoretical (thin lines) results for flat Pb islands of various sizes on different substrates: {\bf a)} Cu(111), {\bf b)} Si(111)-7x7, {\bf c)} HOPG, {\bf d)} 1 ML of h-BN on Ni(111), {\bf e)} 2 MLs of NaCl on Ag(111), and {\bf f)} 3 MLs of NaCl on Ag(111) (not normalized). The inset of {\bf a} shows schematically the experiment $(1)$ and the corresponding electrical circuit $(2)$ used in {\bf b}-{\bf e}. The tip--island junction is characterized by the tunneling resistance $R_T$ and the capacitance $C_T$. The island--substrate contact is modeled as an ohmic resistor $R$ in parallel with a capacitor $C$. Extracted values are indicated together with the island areas in {\bf b}-{\bf e}. In {\bf f}, the tip-island junction and the island-substrate contact are both modeled as tunnel barriers (see page 4) \cite{Hanna1991}. The residual charge on the island is $Q_0=0.064$e.}
\end{figure*}

Figure~\ref{dIdVspectra} displays a selection of differential conductance spectra measured on individual Pb islands of varying sizes on top of the different substrates. The experimental data are corrected for a background contribution due to quantum-well states \cite{Brun2009,Hong2009} and rescaled by the tunneling resistance $R_T$ between the islands and the STM tip such that the normalized spectra approach unity at large voltages. The temperature was below the critical temperature of bulk lead ($T_c=7.2$ K) and one could expect the Pb islands to display superconducting properties for voltages below the superconducting gap of bulk Pb, $|eV|/2<\Delta = $ 1.3 meV \cite{Brun2009}. However, recent experiments have shown that both $T_c$ and $\Delta$ decrease below their bulk values for small systems \cite{Brun2009, Liu2011} and superconducting gap features give a negligible contribution to the conductance under our experimental conditions. On larger voltage scales the quasi-particle transport between the islands and their normal-state substrates is the same as for islands in their normal-state as we checked by performing experiments at temperatures above $T_c$.

While Fig.~\ref{dIdVspectra}a shows that the differential conductance for the islands on Cu(111) essentially remains flat independently of the island size, we observe in \mbox{Figs.~\ref{dIdVspectra}b-e} a suppression of the differential conductance at low voltages which becomes increasingly prominent as the island size is decreased. We attribute the low-bias suppression to DCB due to the high-impedance electrical contact formed between the islands and the substrates. The normalized differential conductance is reduced below unity when the applied voltage $eV$ is smaller than the charging energy $E_C=e^2/C_{\Sigma}$, where $C_{\Sigma}$ is the capacitance of the system. The suppression (and the charging energy) increases as the islands become smaller.  For the islands on Cu(111), the electrical contact has a very low resistance such that the spectrum essentially is ohmic and in this case we ascribe the small features in the spectra to the reduced electron--phonon scattering of the quantum-well states below the Debye energy $E_D \simeq 10$ meV of Pb  \cite{Wang2009,Brun2009}. In Figs.~\ref{dIdVspectra}e-f we show results for Pb islands on Ag(111) covered with 2 and 3 MLs of insulating material (NaCl) between islands and substrates. As several insulating MLs are introduced the spectra begin to display qualitatively different features as we explain below.

\emph{Theory.}---
To understand quantitatively the measured conductance spectra, we employ the $P(E)$--theory of DCB which explicitly incorporates the impedance of the electrical circuit in calculations of the current \cite{Devoret1990,Girvin1990,Ingold1992,Joyez1997b}. The total impedance (as seen from the tunnel junction) reads $Z(\om)=[i\om C_{\Sigma} + Z_{\rm ex}^{-1}(\om)]^{-1}$, where $C_T$ is the capacitance of the tunnel (tip--island) junction and $Z_{\rm ex}(\om)$ is the impedance of the external circuit \cite{Ingold1992}. In our experiment, see inset of Fig.~\ref{dIdVspectra}a, the impedance of the island--substrate contact is $Z_{\rm ex}(\om)=1/(i \om C + R^{-1})$, where $C$ ($R$) is the capacitance (ohmic resistance) of the island--substrate contact, and we obtain $Z(\om)=[i\om C_{\Sigma} + 1/R]^{-1}$ with  $C_{\Sigma}=C+C_T$. The tunneling resistance $R_T=10$~M$\Omega$~--~1~G$\Omega$ between STM tip and island is much larger than the resistance quantum $R_K=h/e^2\simeq 25.8$ k$\Omega$, which justifies a perturbative calculation in the tunnel coupling. The DC current then reads
\begin{equation}
I(V)=(-e)[\Gam_{\mr{tip}\to \mr{isl}}(V)-\Gam_{\mr{isl}\to \mr{tip}}(V)]
\label{eq:I}
\end{equation}
with $\Gam_{\mr{isl}\to \mr{tip}}(V)= \Gam_{\mr{tip}\to \mr{isl}}(-V)$. The tunneling rates are
\begin{equation}
\Gam_{\mr{isl}\to \mr{tip}}(V)=\frac{1}{e^2R_T}\int_{-\infty}^\infty dE \frac{EP(eV+E)}{\exp(E/k_BT)-1},
\label{eq:G}
\end{equation}
where $P(E)$ is the probability for an electron to emit the energy $E$ into the electrical circuit \cite{Ingold1992}. The $P(E)$--function can be written
\begin{equation}
P(E)=\frac{1}{2\pi\hbar}\int_{-\infty}^\infty dt \exp[J(t)+iEt/\hbar],
\label{eq:P}
\end{equation}
where $J(t)=\ew{[{\vp}(t)-{\vp}(0)]{\vp}(0)}$ is the equilibrium correlation function of the phase $\vp(t)=(e/\hbar)\int_{-\infty}^t  dt' V_T(t')$ of the voltage $V_T$ across the tip--island tunnel junction. It can be expressed via the total impedance $Z(\om)$ as
\begin{equation}
J(t)=2\int_0^\infty \frac{d\om}{\om}\frac{\RE Z(\om)}{R_K}\frac{e^{-i\om t}-1}{1-e^{-\hbar\om/k_BT}}
\label{eq:J}
\end{equation}
which we evaluate analytically \cite{Joyez1997}. Finally, we convolve the calculated current with the instrumental resolution function $g_m(\epsilon)=2\Theta(V_{\rm pp}-|\epsilon|)\sqrt{V_{\rm pp}^2-\epsilon^2}/\pi V_{\rm pp}^2$ that accounts for the broadening due to the modulation voltage \cite{Klein1973,Li1998}. The resulting current is then $I_m(V) = \int d\epsilon I(V+\epsilon)g_m(\epsilon)$.

\emph{Results.}---
Figures \ref{dIdVspectra}b-e show calculations of the differential conductance based on Eqs.~(\ref{eq:I}--\ref{eq:J}). The fitting parameters $C$ and $R$ used for the calculations were independently extracted from the experimental data: The charging energy $E_C$ is given mainly  by $C\gg C_T\lesssim$ 1 aF, which determines the width of the differential conductance suppression. The resistance $R$ determines the shape of the curves at small voltages and is adjusted so that the theoretical curves best fit the experimental data.

\begin{figure}
\includegraphics[width=0.8\columnwidth]{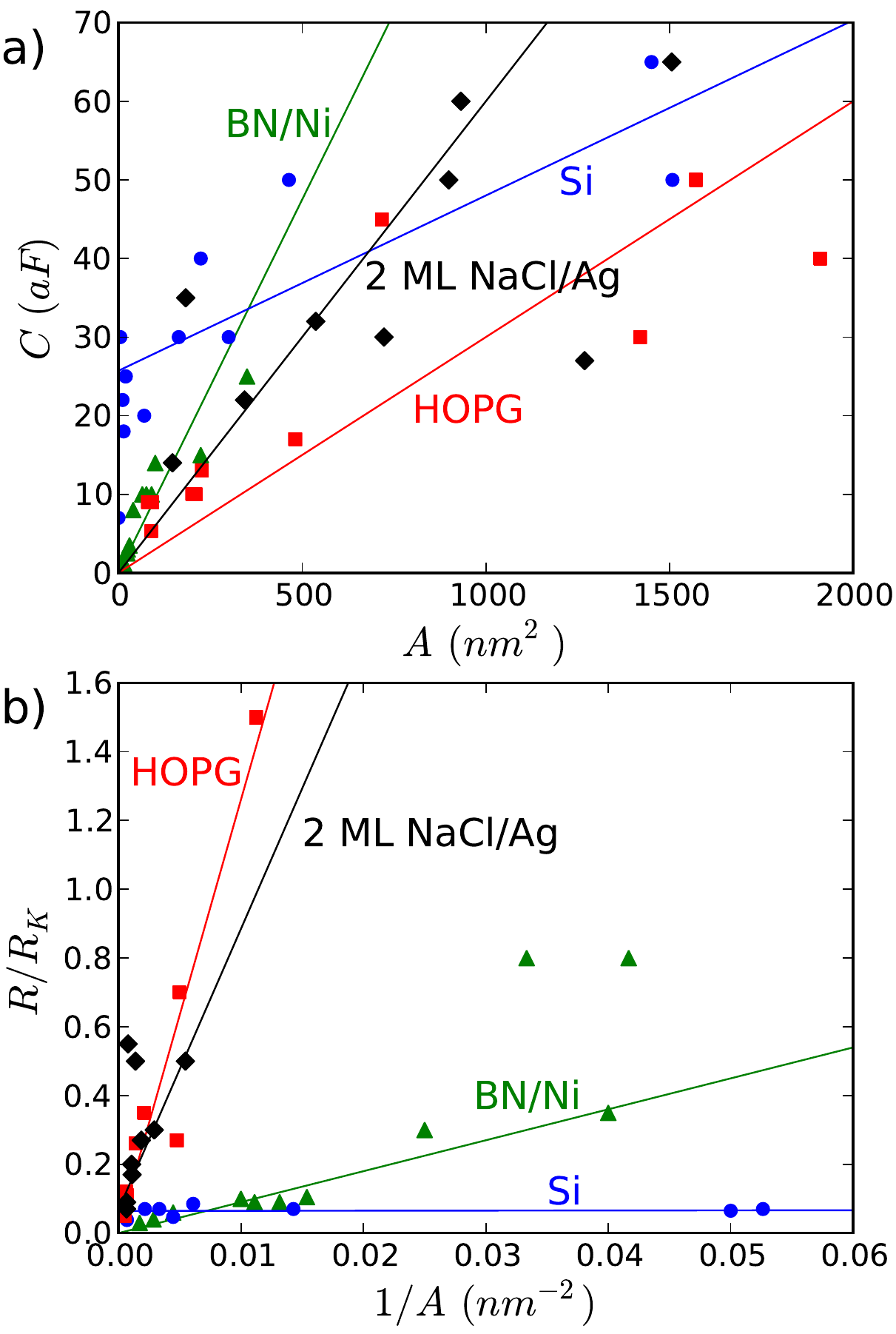}
\caption{\label{fig:results} (color online). Extracted capacitances and resistances. Circles: Pb/Si(111). Squares: Pb/HOPG. Triangles: Pb/BN/Ni(111) Diamonds: Pb/2 ML NaCl/Ag(111). {\bf a)}  Capacitances as functions of the island area $A$. {\bf b)} Resistances as functions of the inverse island area $1/A$. Solid lines are guides to the eye.}
\end{figure}

The theoretical curves are in good agreement with the experimental data and show a clear dependence of the electrical island--substrate contact on the island area. To corroborate our analysis we consider the extracted capacitances and resistances as functions of the island areas. One would expect that the capacitance (resistance) increases with the (inverse) contact area $A$. This systematic behavior is confirmed by Fig.~\ref{fig:results}, showing that we indeed are probing the electrical contact between the islands and the substrates. For HOPG, BN/Ni(111), and 2 ML NaCl/Ag(111) the capacitance (resistance) depends approximately linearly on the (inverse) island area. For Si(111), in contrast, a different behavior is observed: the capacitance has a clear off-set value for small islands and the resistance is essentially independent of the island area. This is due to the wetting layer whose resistance (in parallel) mainly determines the current to the electrical drain contact and whose capacitance does not depend on the island size, giving rise to the off-set at $A\simeq 0$.

Finally, we turn to the samples with several insulating MLs between islands and substrate, Figs.\ \ref{dIdVspectra}e-f. As additional MLs are introduced, electron transport between islands and substrate takes place by tunneling through the insulating layers. The islands are then connected both to the tip and the substrate via tunneling barriers and the orthodox theory of tunneling through a \emph{double} junction  applies \cite{Averin&Likharev}. According to our analysis this occurs with 3 or more MLs of NaCl. In Fig.\ \ref{dIdVspectra}e (2 MLs) some deviations between experiment and the theory of DCB are already visible for large islands and in Fig.\ \ref{dIdVspectra}f (3 MLs) we calculated the spectra using the orthodox theory \cite{Hanna1991}. The gap is associated with the island--substrate junction while the two peaks represent spectral features due to the tip--island junction. The asymmetric gaps in Figs.\ \ref{dIdVspectra}e,f are due to the fractional residual charge $Q_0$ on the Pb islands, which shifts the spectra \cite{Hanna1991}. The controlled addition of single insulating MLs opens an interesting approach to systematic investigations of asymmetric double junctions, similar to recent works on nano-particles coupled to metallic electrodes \cite{Bitton2011,Xu2011}.

\emph{Conclusions.}---
We have used DCB effects to characterize the electrical contact between metallic nano-islands and their supporting substrates in low-temperature STM measurements. Our analysis is supported by the systematic area-dependence of the capacitances and resistances. The present work facilitates quantitative investigations of electrical nano-contacts and is important for future studies of the physical and chemical properties of supported nano-structures in relation to superconductivity, magnetism, and catalysis.

\emph{Acknowledgements.}---
We thank  J.\ P.\ Pekola for indispensable advice and M.\ B\"{u}ttiker, T.\ Cren, and D.\ Roditchev for instructive discussions.
The work was supported by the Swiss National Science Foundation.


\begin{thebibliography}{99}
\expandafter\ifx\csname natexlab\endcsname\relax\def\natexlab#1{#1}\fi
\expandafter\ifx\csname bibnamefont\endcsname\relax
  \def\bibnamefont#1{#1}\fi
\expandafter\ifx\csname bibfnamefont\endcsname\relax
  \def\bibfnamefont#1{#1}\fi
\expandafter\ifx\csname citenamefont\endcsname\relax
  \def\citenamefont#1{#1}\fi
\expandafter\ifx\csname url\endcsname\relax
  \def\url#1{\texttt{#1}}\fi
\expandafter\ifx\csname urlprefix\endcsname\relax\def\urlprefix{URL }\fi
\providecommand{\bibinfo}[2]{#2}
\providecommand{\eprint}[2][]{\url{#2}}

\bibitem[{\citenamefont{Seneor et~al.}(2007)\citenamefont{Seneor,
  Bernand-Mantel, and Petroff}}]{seneor2007}
\bibinfo{author}{\bibfnamefont{P.}~\bibnamefont{Seneor}},
  \bibinfo{author}{\bibfnamefont{A.}~\bibnamefont{Bernand-Mantel}},
  \bibnamefont{and} \bibinfo{author}{\bibfnamefont{F.}~\bibnamefont{Petroff}},
  \bibinfo{journal}{J. Phys. Condens. Matter} \textbf{\bibinfo{volume}{19}},
  \bibinfo{pages}{165222} (\bibinfo{year}{2007}).

\bibitem[{\citenamefont{Bose~\emph{et al.}}(2010)}]{bose2010}
\bibinfo{author}{\bibfnamefont{S.}~\bibnamefont{Bose~\emph{et al.}}},
  \bibinfo{journal}{Nature Mater.} \textbf{\bibinfo{volume}{9}},
  \bibinfo{pages}{550} (\bibinfo{year}{2010}).

\bibitem[{\citenamefont{Lu and Lieber}(2006)}]{Lu2006}
\bibinfo{author}{\bibfnamefont{W.}~\bibnamefont{Lu}} \bibnamefont{and}
  \bibinfo{author}{\bibfnamefont{C.}~\bibnamefont{Lieber}},
  \bibinfo{journal}{Nature} \textbf{\bibinfo{volume}{32}}, \bibinfo{pages}{841}
  (\bibinfo{year}{2006}).

\bibitem[{\citenamefont{Likharev}(2003)}]{Likharev2003}
\bibinfo{author}{\bibfnamefont{K.~K.} \bibnamefont{Likharev}} \bibnamefont{in}
  \emph{\bibinfo{title}{Nano and Giga Challenges in Microelectronics}}
  (\bibinfo{publisher}{Elsevier, Amsterdam}, \bibinfo{year}{2003}).

\bibitem[{\citenamefont{Devoret~\emph{et al}}(1990)}]{Devoret1990}
\bibinfo{author}{\bibfnamefont{M.~H.} \bibnamefont{Devoret~\emph{et al.}}},
  \bibinfo{journal}{Phys. Rev. Lett.} \textbf{\bibinfo{volume}{64}},
  \bibinfo{pages}{1824} (\bibinfo{year}{1990}).

\bibitem[{\citenamefont{Girvin et~al.}(1990)\citenamefont{Girvin, Glazman,
  Jonson, Penn, and Stiles}}]{Girvin1990}
\bibinfo{author}{\bibfnamefont{S.~M.} \bibnamefont{Girvin~\emph{et al.}}},
  \bibinfo{journal}{Phys. Rev. Lett.} \textbf{\bibinfo{volume}{64}},
  \bibinfo{pages}{3183} (\bibinfo{year}{1990}).

\bibitem[{\citenamefont{Ingold and Nazarov}(1992)}]{Ingold1992}
\bibinfo{author}{\bibfnamefont{G.-L.} \bibnamefont{Ingold}} \bibnamefont{and}
  \bibinfo{author}{\bibfnamefont{Yu.~V.} \bibnamefont{Nazarov}} \bibnamefont{in}
  \emph{\bibinfo{title}{Single Charge Tunneling}}
  (\bibinfo{publisher}{Plenum, New York}, \bibinfo{year}{1992}).

\bibitem[{\citenamefont{Cleland~\emph{et al.}}(1992)}]{Cleland1992}
\bibinfo{author}{\bibfnamefont{A.~N.} \bibnamefont{Cleland~\emph{et al.}}},
  \bibinfo{journal}{Phys. Rev. B} \textbf{\bibinfo{volume}{45}},
  \bibinfo{pages}{2950} (\bibinfo{year}{1992}).

\bibitem[{\citenamefont{Holst et~al.}(1994)\citenamefont{Holst, Esteve, Urbina,
  and Devoret}}]{Holst1994}
\bibinfo{author}{\bibfnamefont{T.}~\bibnamefont{Holst}},
  \bibinfo{author}{\bibfnamefont{D.}~\bibnamefont{Esteve}},
  \bibinfo{author}{\bibfnamefont{C.}~\bibnamefont{Urbina}}, \bibnamefont{and}
  \bibinfo{author}{\bibfnamefont{M.~H.} \bibnamefont{Devoret}},
  \bibinfo{journal}{Phys. Rev. Lett.} \textbf{\bibinfo{volume}{73}},
  \bibinfo{pages}{3455} (\bibinfo{year}{1994}).

\bibitem[{\citenamefont{Joyez~\emph{et al.}}(1997)}]{Joyez1997}
\bibinfo{author}{\bibfnamefont{P.}~\bibnamefont{Joyez~\emph{et al.}}},
  \bibinfo{journal}{Phys. Rev. Lett.} \textbf{\bibinfo{volume}{79}},
  \bibinfo{pages}{1349} (\bibinfo{year}{1997}).

\bibitem[{\citenamefont{Zheng~\emph{et al.}}(1998)}]{Zheng1998}
\bibinfo{author}{\bibfnamefont{W.}~\bibnamefont{Zheng~\emph{et al.}}},
  \bibinfo{journal}{Solid State Commun.} \textbf{\bibinfo{volume}{108}},
  \bibinfo{pages}{839} (\bibinfo{year}{1998}).

\bibitem[{\citenamefont{Pierre et~al.}(2001)\citenamefont{Pierre, Pothier,
  Joyez, Birge, Esteve, and Devoret}}]{Pierre2001}
\bibinfo{author}{\bibfnamefont{F.}~\bibnamefont{Pierre~\emph{et al.}}},
  \bibinfo{journal}{Phys. Rev. Lett.} \textbf{\bibinfo{volume}{86}},
  \bibinfo{pages}{1590} (\bibinfo{year}{2001}).

\bibitem[{\citenamefont{Pekola et~al.}(2010)\citenamefont{Pekola, Maisi,
  Kafanov, Chekurov, Kemppinen, Pashkin, Saira, M\"ott\"onen, and
  Tsai}}]{Pekola2010}
\bibinfo{author}{\bibfnamefont{J.~P.} \bibnamefont{Pekola~\emph{et al.}}},
  \bibinfo{journal}{Phys. Rev. Lett.} \textbf{\bibinfo{volume}{105}},
  \bibinfo{pages}{026803} (\bibinfo{year}{2010}).

\bibitem[{\citenamefont{Saira et~al.}(2010)\citenamefont{Saira, M\"ott\"onen,
  Maisi, and Pekola}}]{Saira2010}
\bibinfo{author}{\bibfnamefont{O.-P.} \bibnamefont{Saira}},
  \bibinfo{author}{\bibfnamefont{M.}~\bibnamefont{M\"ott\"onen}},
  \bibinfo{author}{\bibfnamefont{V.~F.} \bibnamefont{Maisi}}, \bibnamefont{and}
  \bibinfo{author}{\bibfnamefont{J.~P.} \bibnamefont{Pekola}},
  \bibinfo{journal}{Phys. Rev. B} \textbf{\bibinfo{volume}{82}},
  \bibinfo{pages}{155443} (\bibinfo{year}{2010}).

\bibitem[{\citenamefont{Parmentier~\emph{et al.}}(2011)}]{Parmentier2011}
\bibinfo{author}{\bibfnamefont{F.~D.}~\bibnamefont{Parmentier~\emph{et al.}}}, \bibinfo{pages}{doi:10.1038/nphys209}.

\bibitem[{\citenamefont{Nagashima~\emph{et al.}}(1995)}]{Nagashima1995}
\bibinfo{author}{\bibfnamefont{A.}~\bibnamefont{Nagashima~\emph{et al.}}},
  \bibinfo{journal}{Phys. Rev. B} \textbf{\bibinfo{volume}{51}},
  \bibinfo{pages}{4606} (\bibinfo{year}{1995}).

\bibitem[{\citenamefont{Repp~\emph{et al.}}(1995)}]{Repp2006}
\bibinfo{author}{\bibfnamefont{J.}~\bibnamefont{Repp}} \bibnamefont{and}
  \bibinfo{author}{\bibfnamefont{G.}~\bibnamefont{Meyer}}
  \bibinfo{journal}{Appl. Phys. A } \textbf{\bibinfo{volume}{85}},
  \bibinfo{pages}{399} (\bibinfo{year}{2006}).

\bibitem[{\citenamefont{Jalochowski~\emph{et al.}}(1988)}]{Jalochowski1988}
\bibinfo{author}{\bibfnamefont{M.}~\bibnamefont{Jalochowski~\emph{et al.}}},
  \bibinfo{journal}{Phys. Rev. B} \textbf{\bibinfo{volume}{38}},
  \bibinfo{pages}{5272} (\bibinfo{year}{1988}).

\bibitem[{\citenamefont{Weitering~\emph{et al.}}(1992)}]{Weitering1992}
\bibinfo{author}{\bibfnamefont{H.~H.} \bibnamefont{Weitering~\emph{et al.}}},
  \bibinfo{journal}{Phys. Rev. B} \textbf{\bibinfo{volume}{45}},
  \bibinfo{pages}{5991} (\bibinfo{year}{1992}).

\bibitem[{\citenamefont{Hinch~\emph{et al.}}(1989)}]{Hinch1989}
\bibinfo{author}{\bibfnamefont{B.~J.} \bibnamefont{Hinch~\emph{et al.}}},
  \bibinfo{journal}{Europhys. Lett.} \textbf{\bibinfo{volume}{10}},
  \bibinfo{pages}{341} (\bibinfo{year}{1989}).

\bibitem[{\citenamefont{Dil~\emph{et al.}}(2007)}]{Dil2007}
\bibinfo{author}{\bibfnamefont{J.~H.} \bibnamefont{Dil~\emph{et al.}}},
  \bibinfo{journal}{Phys. Rev. B} \textbf{\bibinfo{volume}{45}},
  \bibinfo{pages}{161401(R)} (\bibinfo{year}{2007}).

\bibitem[{\citenamefont{Feng et~al.}(2004)\citenamefont{Feng, Conrada,
  Tringides, Kim, and Miceli}}]{Feng2004}
\bibinfo{author}{\bibfnamefont{R.}~\bibnamefont{Feng~\emph{et al.}}},
  \bibinfo{journal}{Appl. Phys. Lett.} \textbf{\bibinfo{volume}{85}},
  \bibinfo{pages}{3866} (\bibinfo{year}{2004}).

\bibitem[{\citenamefont{Gaisch~\emph{et al.}}(1992)}]{Gaisch1992}
\bibinfo{author}{\bibfnamefont{R.}~\bibnamefont{Gaisch~\emph{et al.}}},
  \bibinfo{journal}{Ultramicroscopy} \textbf{\bibinfo{volume}{42}},
  \bibinfo{pages}{1621} (\bibinfo{year}{1992}).

\bibitem[{\citenamefont{Brun~\emph{et al.}}(2009)}]{Brun2009}
\bibinfo{author}{\bibfnamefont{C.}~\bibnamefont{Brun~\emph{et al.}}},
  \bibinfo{journal}{Phys. Rev. Lett.} \textbf{\bibinfo{volume}{102}},
  \bibinfo{pages}{207002} (\bibinfo{year}{2009}).

\bibitem[{\citenamefont{Hong~\emph{et al.}}(2009)}]{Hong2009}
\bibinfo{author}{\bibfnamefont{I-Po}~\bibnamefont{Hong~\emph{et al.}}},
  \bibinfo{journal}{Phys. Rev. R} \textbf{\bibinfo{volume}{80}},
  \bibinfo{pages}{081409(R)} (\bibinfo{year}{2009}).

\bibitem[{\citenamefont{Liu}(2011)}]{Liu2011}
\bibinfo{author}{\bibfnamefont{J.} \bibnamefont{Liu~\emph{et al.}}}, \bibinfo{journal}{J.
  Phys.: Condens. Matter} \textbf{\bibinfo{volume}{23}},
  \bibinfo{pages}{265007} (\bibinfo{year}{2011}).

\bibitem[{\citenamefont{Wang~\emph{et al.}}(2009)}]{Wang2009}
\bibinfo{author}{\bibfnamefont{K.}~\bibnamefont{Wang~\emph{et al.}}},
  \bibinfo{journal}{Phys. Rev. Lett.} \textbf{\bibinfo{volume}{102}},
  \bibinfo{pages}{076801} (\bibinfo{year}{2009}).

\bibitem[{\citenamefont{Joyez and Esteve}(1997)}]{Joyez1997b}
\bibinfo{author}{\bibfnamefont{P.}~\bibnamefont{Joyez}} \bibnamefont{and}
  \bibinfo{author}{\bibfnamefont{D.}~\bibnamefont{Esteve}},
  \bibinfo{journal}{Phys. Rev. B} \textbf{\bibinfo{volume}{56}},
  \bibinfo{pages}{1848} (\bibinfo{year}{1997}).

\bibitem[{\citenamefont{Klein et~al.}(1973)\citenamefont{Klein, Leg\'er, Belin,
  D\'efourneau, and Sangster}}]{Klein1973}
\bibinfo{author}{\bibfnamefont{J.}~\bibnamefont{Klein~\emph{et al.}}},
 \bibinfo{journal}{Phys. Rev. B}
  \textbf{\bibinfo{volume}{7}}, \bibinfo{pages}{2336} (\bibinfo{year}{1973}).

\bibitem[{\citenamefont{Li et~al.}(1998)\citenamefont{Li, Schneider, Berndt,
  and Crampin}}]{Li1998}
\bibinfo{author}{\bibfnamefont{J.}~\bibnamefont{Li}},
  \bibinfo{author}{\bibfnamefont{W.-D.} \bibnamefont{Schneider}},
  \bibinfo{author}{\bibfnamefont{R.}~\bibnamefont{Berndt}}, \bibnamefont{and}
  \bibinfo{author}{\bibfnamefont{S.}~\bibnamefont{Crampin}},
  \bibinfo{journal}{Phys. Rev. Lett.} \textbf{\bibinfo{volume}{81}},
  \bibinfo{pages}{4464} (\bibinfo{year}{1998}).

\bibitem[{\citenamefont{Averin and Likharev}(1991)}]{Averin&Likharev}
\bibinfo{author}{\bibfnamefont{D.~V.} \bibnamefont{Averin}} \bibnamefont{and}
  \bibinfo{author}{\bibfnamefont{K.~K.} \bibnamefont{Likharev}},
  \emph{\bibinfo{title}{Mesoscopic Phenomena in Solids}}
  (\bibinfo{publisher}{Elsevier, Amsterdam}, \bibinfo{year}{1991}).

\bibitem[{\citenamefont{Hanna and Tinkham}(1991)}]{Hanna1991}
\bibinfo{author}{\bibfnamefont{A.~E.} \bibnamefont{Hanna}} \bibnamefont{and}
  \bibinfo{author}{\bibfnamefont{M.}~\bibnamefont{Tinkham}},
  \bibinfo{journal}{Phys. Rev. B} \textbf{\bibinfo{volume}{44}},
  \bibinfo{pages}{5919(R)} (\bibinfo{year}{1991}).

\bibitem[{\citenamefont{Bitton et~al.}(2011)\citenamefont{Bitton, Gutman,
  Berkovits, and Frydman}}]{Bitton2011}
\bibinfo{author}{\bibfnamefont{L.}~\bibnamefont{Bitton}},
  \bibinfo{author}{\bibfnamefont{D.~B.} \bibnamefont{Gutman}},
  \bibinfo{author}{\bibfnamefont{R.}~\bibnamefont{Berkovits}},
  \bibnamefont{and} \bibinfo{author}{\bibfnamefont{A.}~\bibnamefont{Frydman}},
  \bibinfo{journal}{Phys. Rev. Lett.} \textbf{\bibinfo{volume}{106}},
  \bibinfo{pages}{016803} (\bibinfo{year}{2011}).

\bibitem[{\citenamefont{{Xu} et~al.}(2011)\citenamefont{{Xu}, {Sun}, {Yan},
  {Yang}, {He}, {Nie}, and {Li}}}]{Xu2011}
\bibinfo{author}{\bibfnamefont{R.}~\bibnamefont{{Xu~\emph{et al.}}}},
  \bibinfo{journal}{arXiv:1106.2777}  (\bibinfo{year}{2011}).

\end{thebibliography}

\end{document}